\documentclass[aps,prd,nofootinbib,twocolumn,longbibliography]{revtex4}
\usepackage{makeidx}
\usepackage{wrapfig}
\usepackage{sidecap}
\usepackage{enumerate}
\usepackage{feyn}
\usepackage[usenames,dvipsnames]{xcolor}
\usepackage{colordvi} % for color text
\usepackage{amsmath,amssymb}
\usepackage{graphicx}
\usepackage[bookmarks=false,pdfstartview=FitH]{hyperref} %make sure it comes last of your loaded packages

\def\n{{\mathbf n}}
\def\v{{\mathbf v}}

\def\id{1\hspace{-.8mm}\mathrm{l}}
\newcommand{\be}{\nopagebreak[3]\begin{equation}}
\newcommand{\ee}{\end{equation}}
\newcommand{\ba}{\nopagebreak[3]\begin{eqnarray}}
\newcommand{\ea}{\end{eqnarray}}

\begin{document}

\title{\large The spin connection of twisted geometry}

    \author{Hal~M.~Haggard}
    \email{haggard@cpt.univ-mrs.fr}
\address{Centre de Physique Th\'eorique de Luminy,\footnote{Unit\'e de recherche (UMR 6207) du CNRS et du Aix-Marseille Universit\'e; affili\'e \`a la FRUMAM (FR 2291).\vskip2pt} Case 907, 13288 Marseille, France}

    \author{Carlo Rovelli}
    \email{rovelli@cpt.univ-mrs.fr}
\address{Centre de Physique Th\'eorique de Luminy,\footnote{Unit\'e de recherche (UMR 6207) du CNRS et du Aix-Marseille Universit\'e; affili\'e \`a la FRUMAM (FR 2291).\vskip2pt} Case 907, 13288 Marseille, France}

\author{Francesca Vidotto}%
\email{fvidotto@science.ru.nl}
\affiliation{%
Institute for Mathematics, Astrophysics and Particle Physics, 
Radboud University\\
Faculty of Science,  Mailbox 79,  P.O. Box 9010,
6500 GL Nijmegen, The Netherlands
}
        
    \author{Wolfgang Wieland}
    \email{wieland@cpt.univ-mrs.fr}
\address{Centre de Physique Th\'eorique de Luminy,\footnote{Unit\'e de recherche (UMR 6207) du CNRS et du Aix-Marseille Universit\'e; affili\'e \`a la FRUMAM (FR 2291).\vskip2pt} Case 907, 13288 Marseille, France}

\date{\today}

\begin{abstract}
\noindent Twisted geometry is a piecewise-flat  geometry less rigid than Regge geometry. In Loop Gravity, it  provides  the classical limit for each step of the truncation utilized in the definition of the quantum theory. We define the torsionless spin-connection of a twisted geometry. The difficulty given by the discontinuity of the triad is addressed by interpolating between triads. The curvature of the resulting spin connection  reduces to the Regge curvature in the case of a Regge geometry. 
\end{abstract}

%\pacs{04.70.Dy, 04.60.-m}
%\keywords{Twisted Geometries, Quantum gravity}%Use showkeys class option if keyword

\maketitle

\section{Introduction}

Twisted geometry  \cite{Freidel:2010bw,Freidel:2010aq,Dupuis:2012yw,Speziale:2012nu} is a discrete (piecewise-flat) geometry found in loop gravity. Here we define and compute the torsionless spin connection of a twisted geometry. 

In loop gravity, the quantities determining the 3d geometry of physical space are {non-commuting} quantum operators \cite{Rovelli:1994ge,Ashtekar:1998ak},  therefore a quantum geometry is never a \emph{classical} geometry (discrete, twisted, polymeric or otherwise): no more than a quantum particle with spin is a classical rotating sphere.   But the notion of twisted geometry is nevertheless a powerful tool, because it provides the classical limit for each step in the truncation utilized in the definition of the quantum theory \cite{Rovelli:2010km,Rovelli:2011eq}. It is therefore similar to the picture of the states of a quantum field theory as configurations of $n$ classical particles.  

The basic operators of the loop theory are the flux operators, which define the 3d geometry, and the holonomy operators, which define an $SU(2)$ connection on the  same 3d space \cite{Rovelli:2004fk,ThiemannBook,Ashtekar:2004eh}.  Since the two are independent, the connection in general has torsion, as is the case in the continuous (Ashtekar-Barbero) Hamiltonian theory: $SU(2)$-connection degrees of freedom are independent from the 3d-metric degrees of freedom. The mismatch between this connection and the spin connection determined by the intrinsic geometry (namely, by definition, the torsion) codes the information  about the \emph{extrinsic} curvature, which is the canonical variable conjugate to the intrinsic 3-geometry. 

In the continuum theory, the $SU(2)$ connection $A$ is neatly formed by two parts: $A=\Gamma+\gamma K$, where $\gamma$ is the Barbero-Immirzi parameter, $K$ the extrinsic curvature and $\Gamma=\Gamma(e)$ is the spin connection determined by the triad $e$ via the first Cartan structure equation, namely the condition of vanishing torsion.  As pointed out in \cite{Speziale:2012nu}, the same decomposition is not easily achieved in the discrete setting, because the Cartan equation does not make sense on the boundary between piecewise-flat cells.  Therefore a twisted geometry is a generalization of a 3d metric space for which a notion of spin connection has not yet been given. This has made the separation of the part of the connection that codes the extrinsic curvature from the part that doesn't problematic.  In this paper, we fill this gap, providing a definition of $\Gamma(e)$ that remains meaningful in twisted geometry.  
 \begin{figure}[b]
  \includegraphics[scale=0.5]{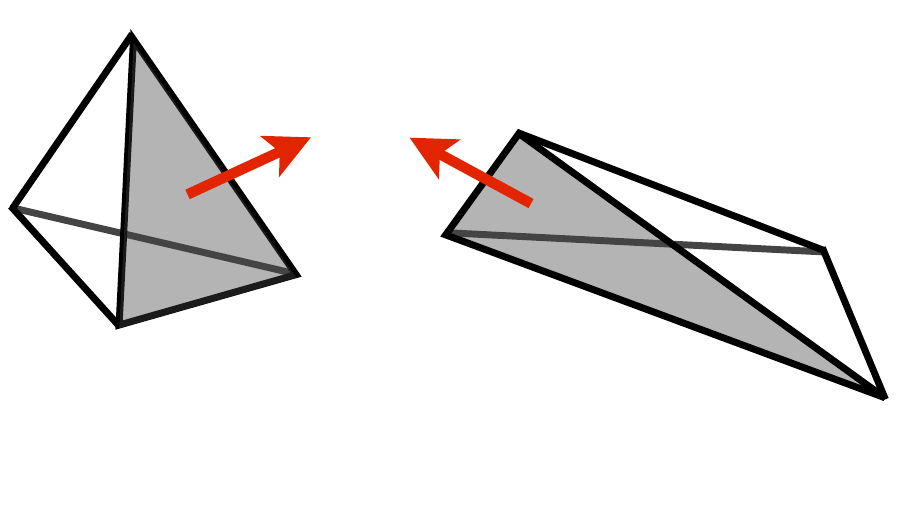}
%\begin{SCfigure}[1.3][h]
%\includegraphics[scale=0.35]{tets.pdf}
\vspace{-6mm}
\caption{In a twisted geometry two adjacent triangles have the same area and the same normal, but the angles and the edge lengths can differ. The two triangles can be identified, at the price of a discontinuity of the metric across the triangle.}
\label{prova}
%\end{SCfigure}
\end{figure}
\section{Definition of the connection}

By a twisted geometry we mean: an oriented 3d simplicial complex (a triangulation) $\cal T$, equipped with a flat metric on each 3-simplex (which makes it a flat tetrahedron), along with the condition that for any two tetrahedra sharing a face the area of the face is the same whether it is computed from the metric on one side or the other (\,FIG. 1\,).  If in addition we require the length of the edges to be the same, we have a Regge geometry. If not, we have a non-Regge twisted geometry.\footnote{This definition is slightly stronger than the one emerging from the classical limit of loop quantum gravity, since it fixes the full triangulation and not just its dual graph. Also, the definition given here refers only to the \emph{intrinsic} geometry. The full definition of the twisted geometry that appears in quantum gravity includes also the \emph{extrinsic} curvature, which plays no role here.  Finally, for simplicity we restrict our attention to triangulations, but the results presented extend to generic cellular decompositions (and therefore to polyhedra other than tetrahedra).}

In general, the metric is discontinuous across a triangle $\tau$. Therefore there are two distinct flat metrics induced on the same face: one from the left tetrahedron and one from the right tetrahedron. The twisting of the geometry measures the difference between these two metrics.  Since a 2d flat metric is determined by three numbers, and, by definition, the two metrics define the same area, there are two twisting parameters. 

Let us setup a coordinate system $(x,y,z)$ covering the two tetrahedra bounding a triangle face $\tau$. It is convenient to choose these coordinates so that the triangle  $\tau$ is at  $z=0$.  Without loss of generality, we can always choose the coordinate system and the triad in such a way that the triad is cartesian, namely $e^i=dx^i$, on the left tetrahedron.  Then the discontinuity of the metric implies that the triad on the right hand side tetrahedron can be chosen to have the constant form 
\be
   e^1=e^1_xdx+e^1_ydy, \ \  \ e^2=e^2_xdx+e^2_ydy,  \  \ \         e^3=dz, 
\ee
The condition that the area is the same from both sides gives $\det e =1$. Therefore the matrix 
\be
e=\{e^i{}_a\}=\left(
\begin{array}{ccc}
 e^1_x& e^1_y  & 0 \\
  e^2_x & e^2_y  & 0  \\
 0 &  0 &   1
\end{array}
\right)
\ee 
is in $SL(3,\mathbb{R})$, or, more specifically it is in the $SL(2,\mathbb{R})$ upper block diagonal subgroup of $SL(3,\mathbb{R})$.  

The geometrical interpretation of these groups is straightforward: $e$ is the linear transformation that sends a cartesian triangle with the dimensions given by the left metric into the cartesian triangle with the dimensions given by the right. In other words, $e$ is the linear transformation that makes the two triangles of FIG. 1 match.   Since the triangle is two dimensional, this linear transformation can always be chosen in the $SL(2,\mathbb{R})$ subgroup.  

On a Riemannian space, once we choose a triad field $e^i$, then the torsionless Cartan spin connection is the unique solution of the first Cartan structure equation
\be
         de^i+\epsilon^i{}_{jk}\ \omega^j\wedge e^k=0.
\label{torsion}
\ee
On a twisted geometry this definition does not make sense, because of the discontinuity of the triad  on the triangles that makes $de^i$ ill defined.  To define the connection on the twisted geometry, we therefore extend this equation ``across" the triangle, where $e^i$ is a discontinuous field.  

For this purpose, let us ``thicken" the triangle, in order to smooth-out the discontinuity, replacing the triangle $\tau$ by a foliated 3d region $\tau \times [0,\Delta]$ where $z\in[0,\Delta]$. Now, we can interpolate the triad by $e(z)$, such that $e(0)=\id$ and $e(\Delta)=e$. The (finite) holonomy of the connection across the face, $U(e)$, can be defined as the $\Delta\to0$ limit of the holonomy of the spin connection of $e(z)$, which is calculated across the thickened triangle. There is a highly nontrivial condition on the interpolating triad: the resulting holonomy must transform as an holonomy under a change of frame on either tetrahedron. That is, 
\be
      U(\Lambda_s e\Lambda_t^{-1})= \Lambda_s U(e) \Lambda_t^{-1} \label{condition} 
\ee
for any $\Lambda_s,\Lambda_t\in SO(3)$. An interpolating triad that satisfies this condition can be obtained starting from the polar decomposition of $e$
\be
     e=e^A e^S
\ee
where $A$ is antisymmetric and $S$ is symmetric, by writing 
\be
     e(z)=e^{zA} e^{zS}.
\ee
This defines a \emph{continuous} triad joining the two tetrahedra, differentiable in $(0,\Delta)$.  We can now compute the spin connection of the interpolating region, and take the limit $\Delta\to0$.  This defines a torsionless spin connection on the twisted geometry.  

 \begin{SCfigure}[1.6][h]
% \put(7,30) {\color{red}$\gamma$}
  \includegraphics[scale=0.5]{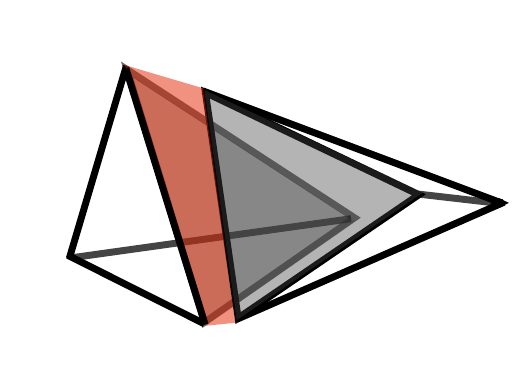}
  \caption{We ``thicken" the triangle in order to smooth-out the discontinuity. The path $\gamma$ goes from one tetrahedron to the other through the thickened region.}
\label{prova}
\end{SCfigure}

Let us compute this connection explicitly. From the last equation, we have 
\be
          de^i=(A+e^{zA}Se^{-zA})^i{}_j\ dz\wedge e^j.
\ee
Using this in the Cartan equation (and lowering an index) we have
\be
          (A+e^{zA}Se^{-zA})_{ij}\ dz\wedge e^j = - \epsilon_{ijk}\ \omega^j \wedge e^k.
\ee
One can check that the solution of this equation is given by 
\be
          \omega^i= B^i{}_j\;  e^j 
\ee
where
\be
          B^i{}_j=-\epsilon^{ikl}(A+e^{zA}Se^{-zA})_{jk}e^z_l +\frac12 \epsilon^{klm}A_{kl}e^z_m\ \delta^i_j,
          \label{term}
\ee
where $e^z_i$ is a matrix element of the inverse triad. 

What is relevant for us is, of course, only the holonomy of this connection across the thickened region.  Consider a path crossing this region at constant $x$ and $y$. The holonomy of $\omega^i$ across the region is given by 
\begin{eqnarray}
  U&=&{\cal P}\, e^{-\int_\gamma \omega}
={\cal P} \,  e^{-\int_0^\Delta  \omega(\partial_z) dz}.
\label{omega1}
\end{eqnarray}
Observe now that since $e(z)\in SL(2,\mathbb{R})\subset SL(3,\mathbb{R})$ it follows that $(A+e^{zA}Se^{-zA})$ is upper block diagonal and so is $B$, and therefore $\omega_z$ is determined just by the second term in \eqref{term}. Explicitly, 
\be
\omega^k(\partial_z)=\frac12 \epsilon^{kij}A_{ij}
\ee
So that 
\be
  U=\exp{A}
\ee
that is, the holonomy is precisely the orthogonal matrix in the polar decomposition of $e$. For the explicit form of the polar decomposition, we have then 
\be
  U(e)=e (e^{\scriptscriptstyle T}e)^{-1/2}
\ee
where $e^{\scriptscriptstyle T}$ is the transpose of $e$. Since $U(e)$ is independent from the size of the interpolating region, taking the limit  $\Delta\to 0$ is immediate.  The resulting distributional torsionless spin connection is concentrated on the face $\tau:(\sigma^1,\sigma^2)\mapsto x^a(\sigma)$ and is given by  
\be
 \Gamma =- A\, d\tau
\ee
where the distributional one-form of the triangle is defined by 
\be
d\tau_a(x) \equiv \int_\tau d^2\sigma \,   \frac{\partial x^b}{\partial\sigma^1}\frac{\partial x^c}{\partial\sigma^2} \, \epsilon_{abc} \, \delta(x-x(\!\sigma\!)).
\ee
It is easy to verify that \eqref{condition} is satisfied.

\section{Connection as a function of the normals}

In this section we compute the connection $U$ in terms of the normals to the faces of the tetrahedra, which are the basic variables defining the twisted geometry in loop gravity.  Let $e$ be a triad in the right tetrahedron and $\tilde e$ the one in the left tetrahedron, in the same coordinate system. The interpolating map is given by the $SL(2,\mathbb{R})$ block diagonal matrix 
\be
   s= e \tilde e^{-1} 
\ee
and the holonomy is $U(s)$.   Consider a tetrahedron defined by the triple of vectors $\v_a\in R^3, a=1,2,3$.  The normals to the faces defined by two of these vectors, normalized to the area of the face, is given by 
\be
    \n_1={\scriptsize \frac12}\; \v_2\times \v_3,
\ee 
and so on cyclically. These equations can be inverted, giving the vectors as functions of the normals:
\be
    \v_1=\frac2{3V}\ \n_2\times \n_3,\label{normals}
\ee 
where $V$ is the volume of the tetrahedron, given by 
\be
V=\frac1{3!}\,(\v_1\times \v_2)\cdot \v_3=\sqrt{\ \frac29\; (\n_1\times \n_2)\cdot \n_3}.
\ee
Say we are interested in the face $f$ defined by the vectors $\v_1$ and $\v_2$, or equivalently by the normal $\n_3$. It is convenient\footnote{Notwithstanding the dimension mismatch.} to use the linear but non-orthogonal coordinates adapted to the face, determined by the triple ${\mathbf  u_a}=(\v_1,\v_2,\hat \n_3)$, where $\hat \n_3=\n_3/|\n_3|$. That is, we use coordinates $x^a=(x,y,z)$ defined by ${\bf x}=x\,\v_1+y\,\v_2+z\,\hat\n_3$. It is immediate to see that in these coordinates the metric of the tetrahedron is given by 
\be
g=\left(
\begin{array}{ccc}
 |\v_1|^2 & ~ \v_1\!\cdot\! \v_2 ~  & 0  \\
  \v_1\!\cdot \!\v_2 & |\v_2|^2  & 0  \\
 0 &  0 &   1
\end{array}
\right)\equiv \left(
\begin{array}{ccc}
 a & ~ b ~  & 0  \\
  c & d  & 0  \\
 0 &  0 &   1
\end{array}
\right).
\ee
Notice that 
{\bf
$|\v_1|^2|\v_2|^2 -  (\v_1\cdot \v_2)^2=(2A)^2$
} (so that $\det g= 4A^2$). 
Without loss of generality, we can orient the cartesian frame (in both the left and  right tetrahedra)  so that 
\begin{eqnarray}
\v_1&=&(a, 0 , 0)\\
\v_2&=&(b, c, 0)\\
\hat\n_3&=&( 0 , 0, 1)
\end{eqnarray}
where
\begin{eqnarray}
a&=& |\v_1|, \ \ \ \
b=\frac{  \v_1\!\cdot\! \v_2}{|\v_1|},\\
c&=&\frac{\sqrt{|\v_1|^2|\v_2|^2-(\v_1\!\cdot\! \v_2)^2}}{|\v_1|}.
\end{eqnarray}
Now, observe that a triad for this metric is precisely  
\be
e^i=v^i_1 dx + v_2^i dy + \hat n_3^i dz,
\ee
that is,
\be
e=\{e^i{}_a\}=\left(
\begin{array}{ccc}
a& 0 & 0  \\
 b & c & 0  \\
 0 & 0 &  1
\end{array}
\right).
\ee
The left triad $\tilde e$ is given by by the same expression for the left tetrahedron, which we indicate here by tilded quantities. 
Therefore the $SL(3,\mathbb{R})$ matrix $s$ that transforms the left triangle into the right one is 
\be
s=\{ e^i{}_a(\tilde e^{-1})^a{}_{j}\}=\left(
\begin{array}{ccc}
  a / \tilde a& 0  & 0  \\
( b \tilde c -c \tilde b  )/\tilde a\tilde c &  c / \tilde c  & 0   \\
 0 &  0 &   1
\end{array}
\right)
\ee
The orthogonal part of the polar decomposition of this matrix is, with some algebra, a rotation in the $xy$ plane with angle determined by
\begin{equation}
\begin{aligned}
\cos(\theta)=(c \tilde{a}+a \tilde{c})/\sqrt{D}, \quad \sin(\theta)=(b\tilde{c}- c\tilde{b})/\sqrt{D},\\
D=\tilde{c}^2 (a^2+b^2)+c^2
   (\tilde{a}^2+\tilde{b}^2)+2 c \tilde{c} (a \tilde{a}-b
   \tilde{b}).
   \end{aligned}
\end{equation}
The holonomy $U$ is a rotation in the plane of the face by this angle, where $a,b,c,\tilde a, \tilde b$ and $\tilde c$ are given explicitly above in terms of the normals.  Finally, the torsionless spin connection is 
\be
  \Gamma= \theta\,  e(\hat\n_3)\,  d\tau.
 \ee
 This gives the torsionless connection explicitly in terms of the normals $\n_i$, which are the independent variables in the loop-gravity twisted-geometry framework.

\section{Curvature}

Let $U_l$ be the holonomy of the connection $\Gamma$ around a circle that surrounds a bone $l$, namely the product of the $U$'s for each tetrahedron meeting at the bone $l$.  Recall that the Regge deficit angle $\delta_l$ of a bone $l$ is defined as $\delta_l=2\pi-\sum_i \theta_i$ where $\theta_i$ are the dihedral angles at $l$ of the $(d-1)$-simplices in the link of $l$.  The following holds:

\emph{Proposition:} If the twisted geometry is Regge, then $U_l$ is a rotation around the axis $e^i(l)$, by an angle equal to the Regge deficit angle. 

To show this, note that the holonomy $U_l$ can always be decomposed into a product of contributions from each tetrahedron meeting at $l$. In turn, the tetrahedral contributions can be further decomposed into a product of two pieces: the holonomy coming from crossing the initial triangle $\tau_i$ upon entering the tetrahedron, $U_{\tau_i}$, and the holonomy arising from changing frames within the tetrahedron $\sigma_i$ in order to adapt to the triangle through which the path leaves the tetrahedron, $U_{\sigma_i}$, thus,
\be
U_{l} = U_{\sigma_n}U_{\tau_{n-1}} \cdots U_{\sigma_1} U_{\tau_1}.
\ee
When the geometry is Regge the triangles all have matching shapes and each of the contributions $U_{\tau_i}$ are the identity. Meanwhile, the changes of frame within each tetrahedron bring the initial triangle's inward normal into the final triangle's outward normal and this is just a rotation about the bone by the dihedral angle, $\theta_i$. Thus the transport around the loop, $U_l$, amounts to rotating the orginal frame by $\delta_l$ just as in Regge calculus. 

Put more simply, the point is that for a Regge geometry the spin connection defined here simply agrees with the spin connection which is defined directly by the fact that there is a flat metric without discontinuities around the bone. This characterization of a Regge geometry is explicit when that geometry is viewed as arising by removing the $(d-2)$-skeleton of a triangulation from a $d$-dimensional manifold $\cal M$ \cite{Bianchi:2009tj}. 

The proposition shows that  in the Regge case the connection defined agrees with the standard torsionless Cartan connection. It is the discrete analog of the relation between the curvature of $\omega(e)$ and the Riemann curvature: if the connection satisfies the Cartan equation, then its curvature $F^{ij}=d\omega^{ij}+\omega^{i}{}_k\wedge\omega^{kj}$ is related to the Riemann tensor of the Riemannian manifold defined by the metric $g_{ab}=e_{ai}e^i_b$ by 
\be
    F^{ij}[\omega(e)]=\frac12 e_{c}^i \, e^{jd}\,  R^c{}_{dab}[g(e)]\ dx^a \wedge dx^b.
\ee
In the general twisted case, the curvature may not be of the characteristic Regge form 
\be
     R_{abcd} \sim e^{\delta  \epsilon_{abe} l^e} \, \epsilon_{cdf} l^f.
\ee
where $\vec l$ is the bone on which the curvature is concentrated. In fact, investigating the general form of the curvature tensor arising from the connection presented here may give insights into the type of generalization that twisted geometries provide. For example, it may be possible to characterize what Petrov classes are possible in a twisted geometry and to see if they are more general than the single class that Regge geometry captures. 

\section{Closing considerations}

We have defined a connection $\Gamma$ in the context of  twisted geometry.  This is determined by the normals to the triangles of the tetrahedra.  It reduces to the standard spin-connection in the Regge case, where its curvature gives the Regge deficit angle.   

The result reinforces the twisted geometry construction, and its interpretation as a classical limit of a truncation of quantum gravity. 

The construction should also contribute to dispelling two possible sources of confusion. The first is the idea that the twisting might code torsion. It does not, since a torsionless connection can be defined in the presence of twisting. The key point is that twisting is a purely metric notion: it refers to discontinuities in the metric, and it is determined by the property of the metric space defined by the discrete geometry.  Torsion, on the other hand, is \emph{not} a purely metric notion: a metric does not define torsion. It is only the existence of a connection \emph{independent} from the metric that can determine a torsion.  Therefore twisting cannot define torsion. The idea of relating twisting and torsion, although intuitively attractive, is misled. 

The second confusion is the idea that twisting needs to be suppressed in order to recover the classical limit of general relativity. A twisted geometry is a generalization of a Regge geometry. It is a discretization of a metric space that is distinct and no less honorable than Regge geometry.  

The conditions under which a twisted geometry reduces to the Regge case have been studied \cite{Dittrich:2008va,Dittrich:2010ey}. Attempts to relate these to the vanishing of the torsion of the four-dimensional spin connection, and therefore to the simplicity constraints of general relativity have been explored \cite{Dittrich:2010ey,Dittrich:2012rj}. But twisting appears in the classical limit of the standard time-gauge Hamiltonian theory, where there are no residual simplicity constraints to deal with. Therefore there is no reason for the simplicity constraints to suppress twisting.  Of course, one can \emph{assume} that the classical limit of discrete general relativity \emph{must} be Regge geometry, but the results presented here put into question the need for this assumption. In particular, there is no clash between the existence of twisting and the possibility of defining the discrete version of the  first Cartan equation.  

Twisted geometry is a bona fide discretization of 3d geometry.

\vskip5mm
\centerline{--------}
\vskip5mm

We thank Simone Speziale for extensive discussions on twisted geometries.
HMH acknowledges support from the
National Science Foundation (NSF) International Research Fellowship Program  (IRFP) under grant OISE-1159218.
FV acknowledges support from the 
Netherlands Organisation for Scientific Research (NWO) Rubicon Fellowship Program.

\vfill

%\newpage
\bibliographystyle{utcaps}
\bibliography{BiblioCarlo, Bib}

\end{document}